# Super high capacity of silicon-carbon anode over 6500 mAh g$^{-1}$ for lithium battery

*Shisheng Lin*[1,2,4*], *Minhui Yang*[1], *Zhuang Zhao*[3,4], *Mingjia Zhi*[3], *Xiaokai Bai*[1]

[1]College of Information Science and Electronic Engineering, Zhejiang University, Hangzhou, 310027, P. R. China.

[2]State Key Laboratory of Modern Optical Instrumentation, Zhejiang University, Hangzhou, 310027, P. R. China.

[3]School of materials science and engineering, Zhejiang University, Hangzhou, 310027, P. R. China.

[4] Zhejiang DiManXi Technology Co., Ltd., Zhejiang, China

*Corresponding author. Email: shishenglin@zju.edu.cn.

## Abstract

As silicon is approaching its theoretical limit for the anode materials in lithium battery, searching for a higher limit is indispensable. Herein, we demonstrate the possibility of achieving ultrahigh capacity over 6500 mAh g$^{-1}$ in silicon-carbon composites. Considering the numerous defects inside the silicon nanostructures, it is deduced the formation of quasi Bose-Einstein condensation should be possible, which can lead to the low viscosity flow of lithium-ions through the anode. At a charge-discharge rate of 0.1C (0.42 A g$^{-1}$), the initial discharge specific capacity reaches 6694.21 mAh g$^{-1}$, with a Coulomb efficiency (CE) of 74.71%, significantly exceeding the theoretical capacity limit of silicon. Further optimization of the anode material ratio results in improved cycling stability, with a discharge specific capacity of 5542.98 mAh g$^{-1}$ and a CE of 85.25% at 0.1C. When the initial discharge capacity is 4043.01 mAh g$^{-1}$, the CE rises to 86.13%. By training a multilayer perceptron with material parameters as inputs and subsequently optimizing it using a constrained genetic algorithm, an initial discharge specific capacity of up to 7789.55 mAh g$^{-1}$ can be achieved theoretically. This study demonstrates that silicon-carbon composites have great potential to significantly enhance the energy density of lithium-ion batteries.

## Introduction

Secondary (also called rechargeable) batteries are the most convenient form of energy storage devices. Among all commercially available secondary batteries, lithium-ion batteries offer the highest energy density, which has a promising future in meeting the requirement for storing energy for vehicles, power grid systems, smartphones and many other daily-life electronics[1]. Nevertheless, breakthrough technologies are highly demanded for increasing the energy density demanding for electric or hybrid-electric vehicles. The demand for fast charging with minor capacity fading requires additionally that these anodes allow a reversible high-rate operation[2,3]. The ideal anode of the lithium battery is usually recognized as metal lithium, which can have a high gravimetrical capacity of 3860 mAh g$^{-1}$ [4,5,6]. Compared to lithium, silicon based materials are promising due to their theoretical capacity and low cost. Through alloying reaction with Li$^+$, the final product of lithiated silicon is Li$_{15}$Si$_4$ with a maximum capacity of 3580 mAh g$^{-1}$ [7]. There are many novel designs of silicon anode materials, including the porus, silicon nanowires, nanoparticles or graphene/silicon structures, amorphous silicon[8-14].

However, is the commonly recognized maximum capacity of 3580 mAh g$^{-1}$ or 4200 mAh g$^{-1}$ a right answer for us[15]? We should seek a different point of view for breaking this limitation. For example, the Li diffusivity in Si reported in literature varies from 10$^{-13}$ to 10$^{-19}$ m$^2$/s. The diffusivity was found at the order of 2×10$^{-17}$ m$^2$/s for the lithium poor phase in first lithiation and 2×10$^{-15}$ m$^2$/s for lithium rich phase and in subsequent cycles. The study of lithiation dynamics is important because it affects both stress generation and rate performance of electrodes for lithium-ion batteries. Defects are important for improving the electrochemical performance of batteries as ion diffusion is important for extra storage sites. It is found that the vacancies in Si can improve the lithiation rate[16]. Regarding electrode materials for lithium batteries, planar defects are critical for improving the lithium-ion diffusion. The planar defects can also facilitate the fabrication of power lithium-ion batteries[17,18].

On the other hand, it is rarely recognized that the transport of Li atoms in silicon seems like the movement of exciton in semiconductor, as the Li atoms and Si atoms form the polar intermetallic compounds. If the electron contributed by Li atom and hole contributed by silicon vacancies bound with each other forming bounded excitons, there is a chance of quasi Bose-Einstein condensation (BEC) happening inside the silicon, which promotes the orderly and high-density arrangement of silicon atoms and lithium atoms. Actually, there are many reports of BEC of excitons in semiconductors[19-24].

This study utilizes the chemical vapor deposition (CVD) method to prepare silicon-carbon composite materials with atomic-level integrated structures with numerous defects by optimizing the anodes of lithium-ion batteries performance of the materials. The experimental results show that the material demonstrates significantly superior performance as an anode in lithium-ion batteries: at a charge-discharge rate of 0.1C, the initial discharge specific capacity reaches 6694.2 mAh $g^{-1}$, with a Coulomb efficiency (CE) of 74.71%, far exceeding the theoretical specific capacity limit of silicon; after further optimization of the material ratio, the composite material exhibits good stability during cycle testing, with an initial discharge specific capacity of 5543.0 mAh $g^{-1}$ and a CE of 85.3%. The results of this study not only provide a new approach for the design of anode materials for lithium-ion batteries but also offer a theoretical foundation and technical support for the development of future high-performance batteries. This innovative material not only demonstrates excellent performance in laboratory tests but also holds great potential for practical applications. Additionally, this research provides important insights for the development of anode materials in other types of ion battery systems (such as sodium-ion and magnesium-ion batteries), offering broad application prospects.

## Results

**Electrochemical performance of the super silicon-carbon composite anodes for LIBs.** The electrochemical performance of active materials was thoroughly analyzed and tested by assembling active material electrode sheets with pure metallic lithium into button-type half-cells, in order to assess their application potential in lithium-ion batteries. Figure 1 presents the galvanostatic charge–discharge (GCD) profiles of the super silicon–carbon composite anodes at a rate of 0.1C for the first three cycles, within a voltage window of 0.005–1.5 V. The specific current corresponding to 1C is defined as 4200 mA g$^{-1}$. In Figure 1(a), for sample SOC1022-1, the initial discharge specific capacity is 6694.21 mAh g$^{-1}$, with a CE of 74.71%; the second discharge specific capacity is 4867.27 mAh g$^{-1}$, with a CE of 91.86%; the third discharge specific capacity is 4394.69 mAh g$^{-1}$, with a CE of 92.68%. In Figure 1(b), for sample SOC1022-2, the initial discharge specific capacity is 6561.36 mAh g$^{-1}$, with a CE of 76.23%; the second discharge specific capacity is 5193.35 mAh g$^{-1}$, with a CE of 97.13%; the third discharge specific capacity is 5039.32 mAh g$^{-1}$, with a CE of 98.05%. In Figure 1(c), for sample SOC1022-3, the initial discharge specific capacity is 6369.96 mAh g$^{-1}$, with a CE of 75.05%; the second discharge specific capacity is 4773.08 mAh g$^{-1}$, with a CE of 91.11%; the third discharge specific capacity is 4287.80 mAh g$^{-1}$, with a CE of 91.19%. In Figure 1(d), for sample SOC1205-1, the initial discharge specific capacity is 5947.54 mAh g$^{-1}$, with a CE of 83.42%; the second discharge specific capacity is 4898.25 mAh g$^{-1}$, with a CE of 92.45%; the third discharge specific capacity is 4445.20 mAh g$^{-1}$, with a CE of 91.44%. In Figure 1(e), for sample SOC1205-2, the initial discharge specific capacity is 5542.98 mAh g$^{-1}$, with a CE of 85.25%; the second discharge specific capacity is 4617.55 mAh g$^{-1}$, with a CE of 94.84%; the third discharge specific capacity is 4272.52 mAh g$^{-1}$, with a CE of 95.10%. And in Figure 1(f), for sample SOC1205-3, the initial discharge specific capacity is 5188.26 mAh g$^{-1}$, with a CE of 82.31%; the second discharge specific capacity is 4352.08 mAh g$^{-1}$, with a CE of 97.42%; the third discharge specific capacity is 4247.34 mAh g$^{-1}$, with a CE of 97.94%. The data in Figure 1 shows that the initial discharge specific capacity of our high-performance half cell is between 5000-7000 mAh g$^{-1}$, and the highest can reach 6694.21 mAh g$^{-1}$ (the CE is 74.71%). The corresponding initial CE is also more than 74%, and the highest can reach 85.25% (the initial discharge specific capacity is 5542.98 mAh g$^{-1}$). It is noted that all those discharge specific capacity exceed the theoretical limitation of silicon.

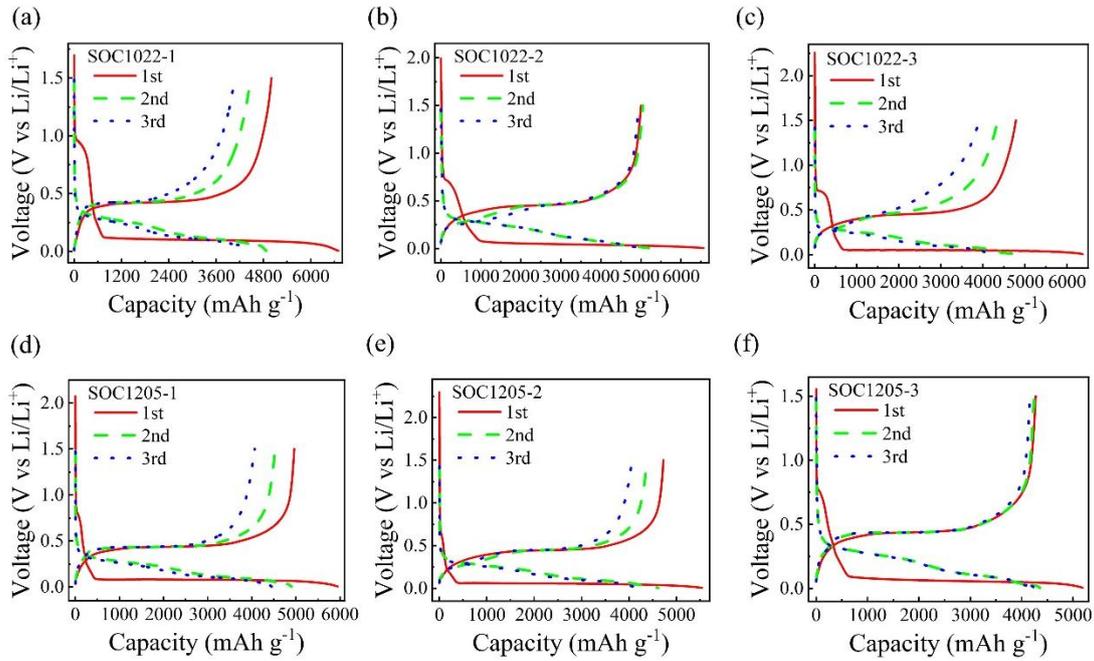

**Fig. 1 | Electrochemical characterization of high specific capacity ratio super silicon-carbon composite anodes upon cycling.** (a) GCD at 0.42 A g$^{-1}$ for the first three cycles for sample SOC1022-1, (b) GCD at 0.42 A g$^{-1}$ for the first three cycles for sample SOC1022-2, (c) GCD at 0.42 A g$^{-1}$ for the first three cycles for sample SOC1022-3, (d) GCD at 0.42 A g$^{-1}$ for the first three cycles for sample SOC1205-1, (e) GCD at 0.42 A g$^{-1}$ for the first three cycles for sample SOC1205-2, (f) GCD at 0.42 A g$^{-1}$ for the first three cycles for sample SOC1205-3.

Figure 2 presents the GCD curves of the super silicon–carbon composite anodes with a high coulombic efficiency ratio, tested at a 0.1C rate over the first three cycles. The voltage window is set from 0.005 V to 1.5 V, with the current density corresponding to 1C defined as 4200 mA g$^{-1}$. In Figure 2(a), for sample SOC1218-1, the initial discharge specific capacity is 4043.01 mAh g$^{-1}$, with a CE of 86.13%; the second discharge specific capacity is 3494.52 mAh g$^{-1}$, with a CE of 95.59%; the third discharge specific capacity is 3295.47 mAh g$^{-1}$, with a CE of 94.88%. In Figure 2(b), for sample SOC1218-2, the initial discharge specific capacity is 3876.75 mAh g$^{-1}$, with a CE of 87.46%; the second discharge specific capacity is 3342.79 mAh g$^{-1}$, with a CE of 95.95%; the third discharge specific capacity is 3148.81 mAh g$^{-1}$, with a CE of 96.37%. In Figure 2(c), for sample SOC1218-3, the initial discharge specific capacity is 3636.44 mAh g$^{-1}$, with a CE of 88.44%; the second discharge specific capacity is 3213.84 mAh g$^{-1}$, with a CE of 97.73%; the third discharge specific capacity is 3128.17 mAh g$^{-1}$, with a CE of 97.84%. In Figure 2(d), for sample SOC1218-4, the initial discharge specific capacity is 3593.73 mAh g$^{-1}$, with a CE of 88.20%; the second discharge specific capacity is 3161.85 mAh g$^{-1}$, with a CE of 97.65%; the third discharge specific capacity is 3076.24 mAh g$^{-1}$, with a CE of 97.75%. In Figure 2(e), for sample SOC1218-5, the initial discharge specific capacity is 3446.41 mAh g$^{-1}$, with a CE of 87.27%; the second discharge specific capacity is 2995.10 mAh g$^{-1}$, with a CE of 97.30%; the third discharge specific capacity is 2872.49 mAh g$^{-1}$, with a CE of 97.70%. In Figure 2(f), for sample SOC1218-6, the initial discharge specific capacity is 3191.80 mAh g$^{-1}$, with a CE of 94.32%; the second discharge specific capacity is 2970.83 mAh g$^{-1}$, with a CE of 94.78%; the third discharge specific capacity is 2751.63 mAh g$^{-1}$, with a CE of 95.78%. The data in Figure 2 shows the charge and discharge data with high initial CE, which ranges from 84 to 95%, and can reach up to 94.32% (the initial discharge specific capacity is 3191.80 mAh g$^{-1}$). The initial discharge specific capacity ranges from 3000 to 5000 mAh g$^{-1}$, and can reach up to 4988.61 mAh g$^{-1}$ (the CE is 84.55%). Table 1 shows the initial efficiency and initial discharge specific capacity of 12 samples in Figures 1 and 2.

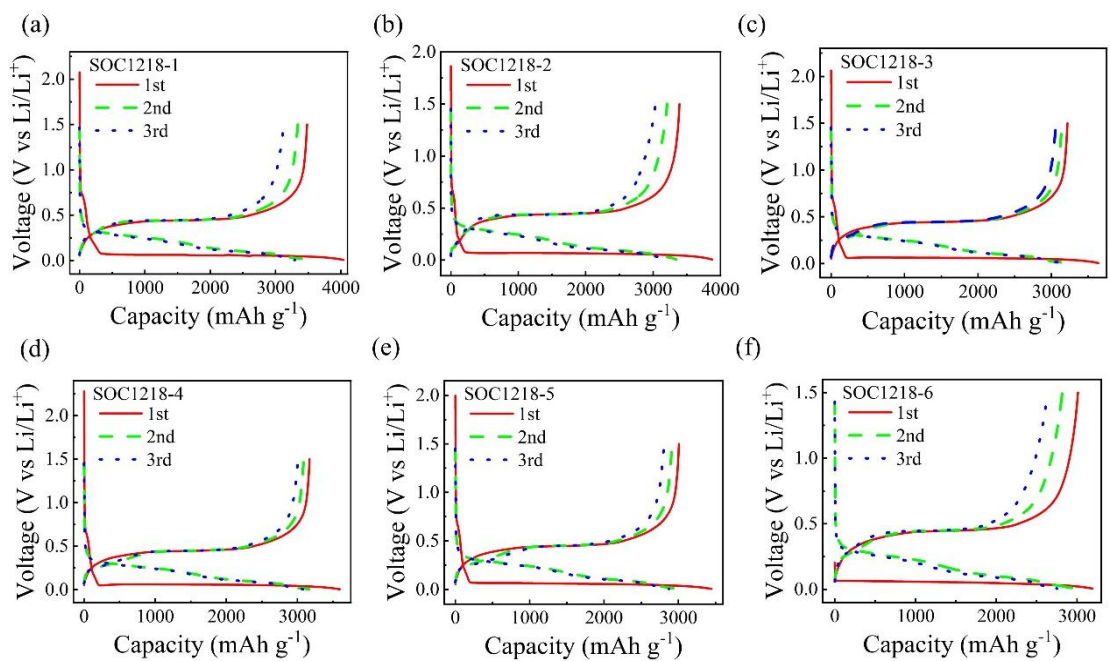

**Fig. 2 | Electrochemical characterization of super silicon-carbon composite anodes upon cycling.** (a) GCD at 0.42 A g$^{-1}$ for the first three cycles for sample for sample SOC1218-1, (b) GCD at 0.42 A g$^{-1}$ for the first three cycles for sample SOC1218-2, (c) GCD at 0.42 A g$^{-1}$ for the first three cycles for sample SOC1218-3, (d) GCD at 0.42 A g$^{-1}$ for the first three cycles for sample SOC1218-4, (e) GCD at 0.42 A g$^{-1}$ for the first three cycles for sample SOC1218-5, (f) GCD at 0.42 A g$^{-1}$ for the first three cycles for sample SOC1218-6.

**Table 1.** The initial discharge specific capacity and Coulomb efficiency of 12 samples in Figures 1 and 2.

| Sample name | The initial discharge specific capacity (mAh g$^{-1}$) | Coulombic efficiency(%) | Sample name | The initial discharge specific capacity (mAh g$^{-1}$) | Coulombic efficiency(%) |
|---|---|---|---|---|---|
| SOC1022-1 | 6694.21 | 74.71 | SOC1218-1 | 4043.01 | 86.13 |
| SOC1022-2 | 6561.36 | 76.23 | SOC1218-2 | 3876.75 | 87.46 |
| SOC1022-3 | 6369.96 | 75.05 | SOC1218-3 | 3636.44 | 88.44 |
| SOC1205-1 | 5947.54 | 83.42 | SOC1218-4 | 3593.73 | 88.20 |
| SOC1205-2 | 5542.98 | 85.25 | SOC1218-5 | 3446.41 | 87.27 |
| SOC1205-3 | 5188.26 | 82.31 | SOC1218-6 | 3191.80 | 94.32 |

It is traditionally assumed that the chemical composition of electrode materials is determined under static equilibrium conditions. However, in electrochemical systems involving mobile lithium-ions, such an assumption may oversimplify the actual dynamics. We propose a dynamic chemical composition model in which not only lithium but also a fraction of silicon atoms within the defective silicon–carbon framework is mobile during electrochemical cycling. This co-migration enables the transient formation of lithium–silicon complexes with polar characteristics. Analogous to BEC observed at ultralow temperatures, these complexes may undergo a form of dynamic, collective transport. We refer to this phenomenon as "quasi" BEC to emphasize the departure from the strict statistical conditions required for conventional BEC, due to disparities in atomic mass and mobility. The significant mass difference between lithium and silicon results in a marked disparity in their diffusion rates, with lithium migrating orders of magnitude faster than silicon. This imbalance leads to a dynamic accumulation of lithium-rich lithium–silicon complexes, producing a time-dependent stoichiometry that deviates from the equilibrium composition. Such a mechanism can account for the anomalously high lithium storage capacity observed in certain super silicon–carbon composite anodes, where the effective lithium content surpasses the theoretical limit inferred from static structural considerations.

Figure 3 shows the TEM image of the super silicon-carbon composite anodes before cycling. As shown in Figure 3(a) and 3(b), many silicon nanodots are supported by the graphene sheet, demonstrating the good interface between silicon and graphene by the CVD method. From Figure 3(c), the graphene and defective silicon lattice can be clearly seen.

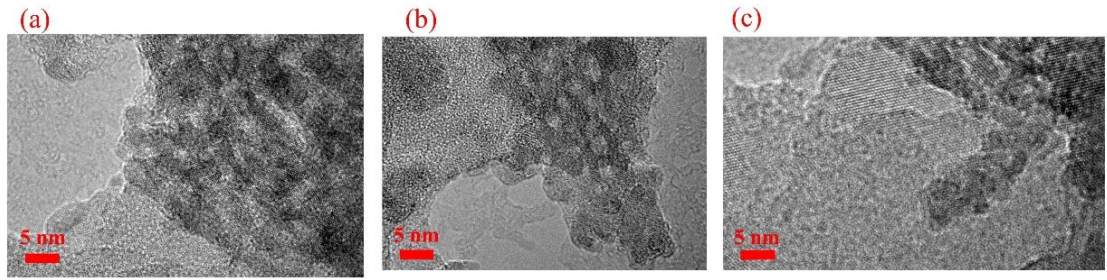

**Fig. 3 | Morphological characterization.** TEM image of graphene supported Si nanodots (a), (b) and HRTEM of graphene supported Si (c).

For predicting the initial CE and the initial discharge specific capacity of super silicon–carbon composite anodes, we developed a multi-layer perceptron (MLP)[25], trained and validated on a dataset of 96 experimentally measured synthesis performance pairs. The input features included five primary compositional variables, corresponding to the concentrations of active silicon-carbon material, carbon black, carbon additive, PVDF binder, and the type of carbon additive. We introduced a second-order feature crossing strategy[26,27] before training to better capture complex interactions among material parameters. Specifically, we computed all pairwise element-wise products—known as the Hadamard products[28]—between the original input features. This operation enriches the input space with nonlinear interaction terms without introducing additional model parameters. Mathematically, given an original input vector $x = [x_1, x_2, \ldots, x_n]$, the Hadamard product generates new features of the form $x_i \cdot x_j$ for all $i < j$, resulting in a total of $n(n-1)/2$ additional features. In our case, the five original features were expanded to include ten additional second-order features, yielding a total of fifteen inputs to the neural network. This augmentation enables the model to learn nonlinear combinatorial effects among the formulation components, which are often critical in governing electrochemical performance, yet are difficult to capture using only first-order descriptors.

The dataset then randomly partitioned into training (70%), validation (15%), and test (15%) subsets. All features and targets were standardized using z-score normalization to facilitate stable and efficient model training. The neural network architecture comprised a feature input layer followed by four fully connected hidden layers with 512, 256, 128, and 64 neurons, respectively as shown in Figure 4(a). The first two layers were regularized using dropout (rates of 30% and 20%) and batch normalization, with Leaky ReLU and ReLU activation functions applied sequentially to introduce non-linearity and avoid vanishing gradient effects. A final regression layer was used to output a continuous prediction of the target property. In addition, a custom warm-up learning rate schedule was implemented to stabilize training using stochastic gradient descent with momentum (SGDM)[29], with a gradually increasing learning rate during the initial epochs. The model was trained for 300 epochs with a mini-batch size of 32 and an initial learning rate of $1\times10^{-2}$. The loss function curve for 600 iterations is shown in the Figure 4(b).

After training, the model demonstrated excellent predictive performance on the test set, with a root-mean-square error (RMSE) of 207.98, mean absolute error (MAE) of 145.14, and a coefficient of determination ($R^2$) of 0.94 for the initial discharge capacity. For the initial CE, the RMSE was 1.25, MAE was 1.08, and $R^2$ reached 0.90. As shown in Figure 4(c), the predicted values of initial discharge capacity correlate strongly with experimental results, with 85.71% of predictions falling within a 5% error margin, enhanced by supplementing the charge and discharge data of the second cycle as input. Figure 4(d) presents a sample-wise comparison for discharge capacity. Meanwhile, Figure 4(e) compares predicted and experimental values for CE, where the model achieves 100% prediction accuracy within a 5% error threshold, highlighting its robustness and consistency. Subsequently, the trained MLP model was integrated with a genetic algorithm (GA)[30] framework to obtain the theoretical maximum initial discharge specific capacity and initial CE. While the MLP provides accurate and efficient predictions of electrochemical properties from given input features, it alone does not offer a mechanism for systematically identifying the optimal combinations of input parameters. GA, a global optimization method inspired by natural evolution, is particularly well-suited for navigating high-dimensional, nonlinear, and non-convex spaces that often characterize materials design problems. By using the MLP as a fast surrogate evaluator within the GA framework, the search process can efficiently converge toward the maximum target properties. This synergistic integration of MLP with GA enables the discovery of theoretical performance limits. The GA was configured with a population size of 100 and a maximum of 200 generations, subject to both linear equality and nonlinear inequality constraints to ensure physically feasible input formulations. Through this approach, the maximum achievable initial CE and discharge specific capacity were predicted to be 88.45% and 7789.55 mAh g$^{-1}$, respectively, underscoring the potential of data-driven inverse design strategies in accelerating battery materials development.

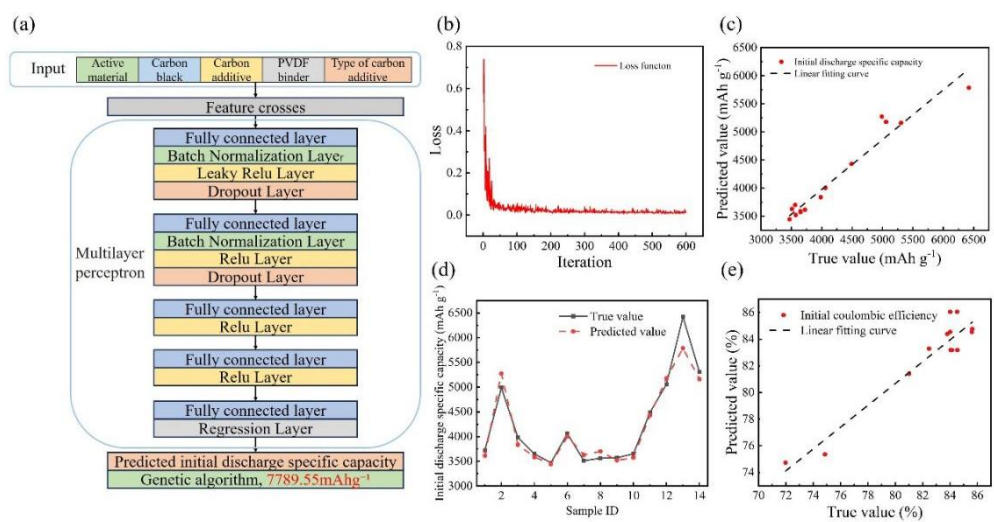

**Fig. 4 | Results of AI prediction.** (a) Diagram of the network structure, (b) Loss function curve, (c) Comparison between predicted and true values of the initial discharge specific capacity, (d) Error analysis for the prediction of the initial discharge specific capacity, (e) Comparison between predicted and true values of the initial CE.

## Conclusion

In summary, we report super silicon–carbon composite anodes with lithium storage capacities exceeding 6500 mAh g$^{-1}$, significantly surpassing the theoretical limit of silicon. This anomalous capacity is attributed to a dynamic chemical composition mechanism, wherein both lithium-ions and a fraction of silicon atoms co-migrate within a defect-rich silicon–carbon framework during cycling. The transient formation of polar lithium–silicon complexes, facilitated by disparities in atomic mass and mobility, leads to a dynamic accumulation of lithium-rich species. We propose a "quasi" BEC transport behavior to describe this collective migration process, offering a physical basis for the time-dependent deviation from equilibrium stoichiometry. Furthermore, a multilayer perceptron combined with genetic algorithm predicts a theoretical maximum initial discharge specific capacity of 7789.55 mAh g$^{-1}$. These findings not only challenge the conventional capacity ceiling of silicon-based anodes but also open new avenues for designing high-performance electrode materials through defect engineering and dynamic stoichiometric ratio.

## Experimental Section

**Characterization Methods.** Scanning electron microscopy (SEM) measurements were collected using a Hitachi S-4800 (10 kV). HAADF images were acquired using aberration corrected STEM (HF5000, Hitachi) operated at the acceleration voltage of 200 kV.

**Electrochemical characterization.** The working electrodes were made by a typical slurry method with active materials (two-dimensional silicon-carbon), conductive additive (Ketjen black), and polyvinylidene fluoride (PVDF) binder with a mass ratio of 8:1:1 were dispersed in N-methyl-2-pyrrolidone (NMP). After casting onto a 15-μm-thick Cu foil and drying at 80 °C in a vacuum oven for 12 h, the samples were cut into 113.04 mm$^2$ circular disks with a mass loading of ~1-2 mg cm$^{-2}$. In an Ar-filled glovebox, these working electrodes were assembled into type 2032 coin cells with the glass microfiber filters separator (Whatman, GF/D) and Li metal as the counter/reference electrode (half cell). Subsequently, 100 μl of 1.0 M LiPF$_6$ in EC:DEC=1:1 Vol% was added as the electrolyte with full wetting of both working and counter electrode surfaces. Electrochemical tests were carried out in the voltage window between 0.005 and 1.5 V.

## Acknowledgements

S. L. thanks the support from the National Natural Science Foundation of China (Nos. 51202216, 51551203, 61774135, and 62474161), the Distinguished Youth Fund of Zhejiang Natural Science Foundation of China (LR21F040001).

## Author Contributions

S. L. designed the experiments, coordinated the experiments, carried out the experiments, discussed the results and wrote the paper. M. Y. and Z. Z. carried out the experiments, discussed the results, and wrote the paper. M. Z. and X. B. discussed the results and assisted with experiments. All authors contributed to the preparation of the manuscript.